\documentclass[showpacs,twocolumn,amsmath,amssymb]{revtex4}

\usepackage{graphicx}
\usepackage{color}
\usepackage{mathptmx}
\usepackage{esvect}

\begin{document}


\title{Effect of Memory on the Dynamics of Random Walks on Networks}


\author{Renaud Lambiotte\email{renaud.lambiotte@unamur.be}}
\affiliation{naXys, University of Namur, Rempart de la Vierge 8, 5000 Namur, Belgium}

\author{Vsevolod Salnikov\email{vsevolod.salnikov@unamur.be}}
\affiliation{naXys, University of Namur, Rempart de la Vierge 8, 5000 Namur, Belgium}

\author{Martin Rosvall}
\affiliation{Integrated Science Lab, Department of Physics, Ume{\aa} University, SE-901 87 Ume{\aa}, Sweden}

\date{\today}

\begin{abstract}
Pathways of diffusion observed in real-world systems often require stochastic processes going beyond first-order Markov models, as implicitly assumed in network theory.
 In this work, we focus on second-order Markov models, and derive an analytical expression for the effect of memory on the spectral gap and thus, equivalently, on the characteristic time needed for the stochastic process to asymptotically reach equilibrium. Perturbation analysis shows that standard first-order Markov models can either overestimate or underestimate the diffusion rate of flows across the modular structure of a system captured by a second-order Markov network. We test the theoretical predictions on a toy example and on numerical data, and discuss their implications for network theory, in particular in the case of temporal or multiplex networks.
\end{abstract}

\pacs{89.75.-k,89.90.+n}

\maketitle

\section{Introduction}
Over the last 15 years, network science has developed a set of powerful tools and models for complex systems made of interacting elements \cite{Boccaletti2006,Newman2010}. The fundamental step, when adopting a network perspective, is the definition of a static underlying medium, made of nodes and links, on which to apply a dynamical process, e.g. synchronization, consensus or diffusion, in order to model the system in question. By implicitly decoupling dynamics and structure \cite{Batty1979,Flow}, this approach allows for the analysis of a broad range of systems within a single framework. However, this abstracting step discards many details from real-life systems, often for good reasons, but it is yet unresolved  where to set the limits of validity of network theory. In this respect, it has recently been shown that flow pathways observed in a broad range of social and information systems, such as web traffic \cite{meiss2008ranking,chierichetti2012web}, human \cite{gonzalez2008understanding} and cattle \cite{heath2008construction} mobility, tend to differ from those reproduced by network-based models. This difference stems from the memory in the pathways, as the direction of a flow strongly depends on where it comes from. In \cite{Rosvall}, Rosvall et al. have shown in a variety of empirical systems that memory constraints on flow are statistically significant, and that a second-order Markov model often provides a good approximation of the dynamics. The authors also showed that memory strongly affects the outcome of algorithms for ranking the importance of nodes or for clustering them, and observed that it has the tendency to slow down spreading, in agreement with previous works \cite{balcan2009multiscale,belik2011natural}.

\begin{figure} 
	\centerline{\includegraphics[width=0.4\textwidth]{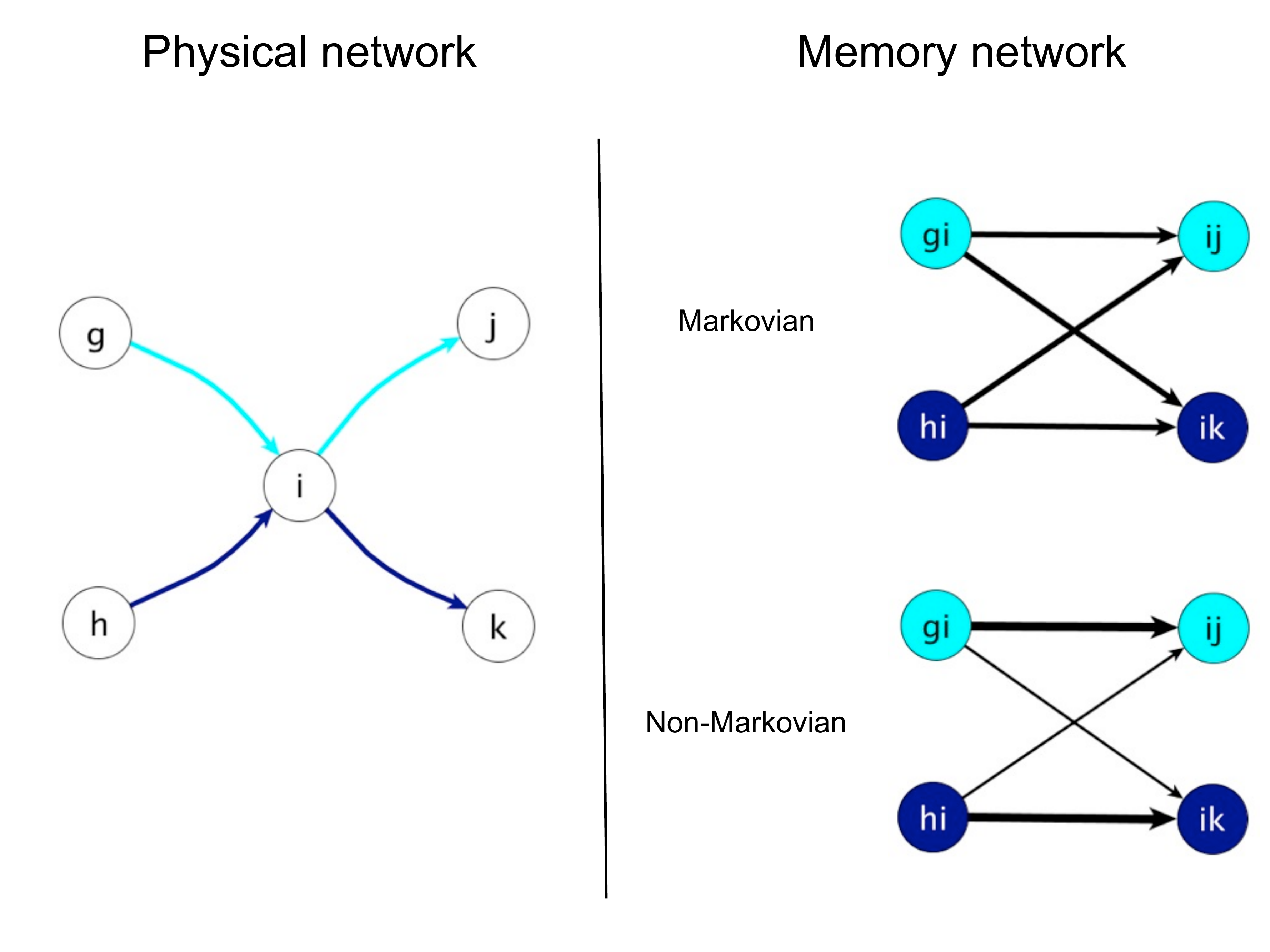}}
	\caption{\label{fig1} 
	Illustration of the effect of memory on patterns of flows. In the left panel, a standard network is used to represent transitions between nodes. In the right panel, the representation is enriched up to second-order Markov, by focusing on the transitions between directed links, i.e. memory nodes.  Keeping the underlying topology fixed, memory networks allow for the modelling of non-Markovian constraints on flow, as can be observed in empirical systems. When the dynamics is memoryless, the transition between connected memory nodes is uniform. 
Non-markovianity leads to biases in the random walk process.} 
\end{figure}

The main purpose of this paper is to derive an analytical expression for the effect of memory on the dynamics of random walkers on networks. To do so, we focus on second-order Markov processes and show the effect of memory on the spectral gap and thus, equivalently, on the mixing time, e.g. the time for the stochastic process to reach equilibrium. The theoretical results are tested numerically on a toy example and on empirical data, and show the conditions under which memory either slows down or accelerate diffusion on networks. Finally, we draw connections with spectral graph theory and community detection, and discuss the general implications of this work for the study of networks.

\section{From first to second order Markov}

For the sake of simplicity, let us first consider an undirected, unweighted and connected network composed of $N$ nodes and $L$ links.
The graph is described by its $N \times N$, symmetric adjacency matrix $A_{ij}$, with $A_{ij}=1$ if there is a link between $i$ and $j$, and zero otherwise. The degree of a node is defined by $k_i=\sum_j A_{ij}$.
The dynamics of a random walker with memory on this physical network is modeled by using the concept of memory network \cite{Rosvall}. The dynamics are encoded by a transition matrix with elements of the form
\begin{align}
\label{aa}
T(\vv{ij} \rightarrow \vv{jk}),
\end{align}
measuring the probability that the walker steps from $j$ to $k$ if it came from $i$ in the previous step, that is from memory node $\vv{ij}$ to memory node $\vv{jk}$.
The matrix is normalized such that $\sum_k T(\vv{ij} \rightarrow \vv{jk}) = 1$. In this framework, the non-Markov process on the physical nodes is modeled by a Markov process on the memory nodes, i.e. as a second-order Markov process. There are $2L$ memory nodes because each undirected link, say between $i$ and $j$ in the physical network leads to 2 memory nodes, $\vv{ij}$ and $\vv{ji}$, in order to encode the time ordering of the visits \footnote{In the case of directed networks, a network of $L$ directed links leads to a memory network of $L$ memory nodes.}. Moreover, the corresponding memory network is directed even if the physical network, as a transition between $\vv{jk}$ and $\vv{ij}$ is forbidden even if a transition between $\vv{ij}$ and $\vv{jk}$ exists.

From now on, let us use Greek letters to denote memory nodes, and Latin letters to denote physical nodes. By definition, the structure of the transition matrix $T_{\alpha \beta} \equiv T(\alpha \rightarrow \beta)$ univocally determines the properties of the stochastic process. 
In the case of a continuous-time random walk process with Poisson waiting times, the probability $P(\alpha;t)$ of finding a walker on memory node $\alpha$ at time $t$ evolves as
\begin{equation}
\dot{P}(\beta;t) =-  \sum_{\alpha} P(\alpha;t) (I_{\alpha \beta} - T_{\alpha \beta}),
\end{equation}
where ${\rm I}$ is the identity matrix and ${\rm L=I-T}$ is the normalized Laplacian of the process.
 If the original process is Markovian, the transition to any out-neighbour of $\alpha$ is equally likely and
\begin{equation}
\label{mar}
T^{M}_{\alpha \beta}= 
\begin{cases}
1/k_\alpha^{\rm out}      & {\rm for}~~ \beta \in \sigma^{\rm out}_\alpha,\cr
0      & {\rm otherwise} ,
\end{cases} 
\end{equation}
where $\sigma^{\rm out}_\alpha$ is the set of out-neighbours of $\alpha$, and $k_\alpha^{\rm out}$ is the size of this set. By definition,  $k_\alpha^{\rm out}=k_j$, where $j$ is the end of memory node $\alpha=\vv{ij}$, because the network is undirected. After defining the probability to find a walker on physical node $j$ at time $t$
\begin{equation}
P(j;t)=\sum_{i} P(\vv{ij};t), 
\end{equation}
it is straightforward  to show that the standard rate equation
\begin{equation}
\label{standardnode}
\dot{P}(j;t) = - \sum_{i} P(i;t) (I_{ij} - \frac{A_{ij} }{k_i}),
\end{equation}
is recovered \cite{Rosvall}.

In general, the stationary state ${\rm \pi }$ is given by the left eigenvector of eigenvalue $0$ of ${\rm L}$
\begin{align}
{\rm \pi L} =0.
\end{align}
In the Markovian case (\ref{mar}), the stationary solution of the process is $\pi(\alpha)=1/2L$ \cite{Rosvall}, in agreement with the fact that the stationary probability of finding a walker on a physical node is proportional to its degree in undirected networks. Since now on, we will assume that the memory network is strongly connected to ensure the unicity of the stationary solution. 

\section{Effect of memory on mixing time}

In this work, we are interested in the transient properties of the diffusive process and, in particular, in the asymptotic relaxation towards equilibrium. This relaxation is determined by the second dominant eigenvalue of L, solution of 
\begin{align}
{\rm u L} = \lambda_2 {\rm u},
\end{align}
where  $\lambda_2$ is the lowest positive eigenvalue of L, i.e. that associated to the slowest mode. The characteristic time to reach equilibrium, called mixing time, is  $1/Re(\lambda_2)$. This eigenvalue, usually called spectral gap, determines the speed of convergence towards stationarity, and is deeply connected to the modular organization of the underlying network \cite{Delvenne}: small values of the spectral gap correspond to networks made of two well-defined communities. 
In the case of undirected networks, the second dominant eigenvector ${\rm u}$, known as the normalized Fiedler vector \cite{Fiedler}, is used in heuristics for graph bi-partitioning \cite{Shi} or modularity optimisation \cite{Delvenne,Laplacian,Newman}.
In particular, the bipartition of a graph is deduced from the sign of entries of ${\rm u}$. Because the transition matrix L is asymmetric, the right eigenvector  
\begin{align}
{\rm L v^{T}} = \lambda_2 {\rm v^{T}}
\end{align}
is in general different from the left eigenvector. Because the underlying network is undirected and assuming that the dynamics is Markovian, a particular structure of memory networks, themselves directed, leads to a connection between their left and right eigenvectors. Let ${\rm G}$ be a memory network and ${\rm G'}$ be the memory network obtained by reversing all links between memory nodes. ${\rm G}$ and ${\rm G'}$ are different, as noted before, but they are isomorphic, through the transformation $\vv{ij} \rightarrow \vv{ji}$ for all memory node \footnote{The detailed spectral properties of memory networks will be the subject of a future work.}.

The effect of memory on $\lambda_2$ is evaluated by perturbation analysis, typically used to determine the effect of structural perturbations on the eigenvalue spectra of networks \cite{Restrepo,Milanese,Masuda}.  Let us consider the Markovian transition matrix (\ref{mar}) as a baseline, and a small deviation around it due to memory 
\begin{align}
{\rm T} = {\rm T^{M}} + {\rm \Delta T}.
\end{align}
It is important to note that ${\rm \Delta T}$ induces a bias in the random walk on the memory network, and that $\Delta T_{\alpha \beta} >0$ ($<0$) for transitions that are promoted (hindered) by memory (see Figure 1). Moreover, conservation of probability implies that $\sum_\beta \Delta T_{\alpha \beta}=0$. By definition, the corresponding Laplacian operator is defined as ${\rm L^M}={\rm I} - {\rm T^M}$, and ${\rm \Delta L}=- {\rm \Delta T}$.
 For the sake of simplicity, we further assume that $\lambda_2$ is not degenerate.
Writing ${\rm u} = {\rm u^M} + {\rm \Delta u}$, ${\rm v} = {\rm v^M} + {\rm \Delta v}$ and $\lambda_2 = \lambda_2^M + \Delta \lambda_2$, and keeping only linear terms, one finds
\begin{eqnarray}
- {\rm  u^{M} \Delta T + \Delta u L^{M} } = \lambda_2^M {\rm \Delta u} + \Delta {\rm\lambda_2 \rm  u^{M} }. 
\end{eqnarray}
The perturbation of the second dominant eigenvalue is obtained by multiplying by the right eigenvector
\begin{eqnarray}
\label{res}
 \Delta \lambda_2 
 = - \frac{ \sum_{\alpha \beta} u^M_\alpha \Delta T_{\alpha \beta} v^M_\beta}{\sum_{\alpha} u^M_\alpha v^M_\alpha}.
\end{eqnarray}

This expression identifies the effect of memory on $\lambda_2$ and on the mixing time of the stochastic process. The spectral gap decreases and the mixing time increases when $\Delta T_{\alpha \beta}$ is positive for nodes in the same community, and negative across communities. In contrast, the spectral gap increases when cross-community flows are favoured by the memory bias. 
The two communities are those defined by the signs of the normalized Fiedler vector, as mentioned before.
These result are intuitive as the network dynamics becomes more (less) encapsulated inside communities, and that the network becomes effectively more (less) modular when cross-community flows decrease (increase). Nonetheless, (\ref{res}) has the advantage of quantifying the effect of memory on spreading.

\begin{figure} 
	\centerline{\includegraphics[width=0.3\textwidth]{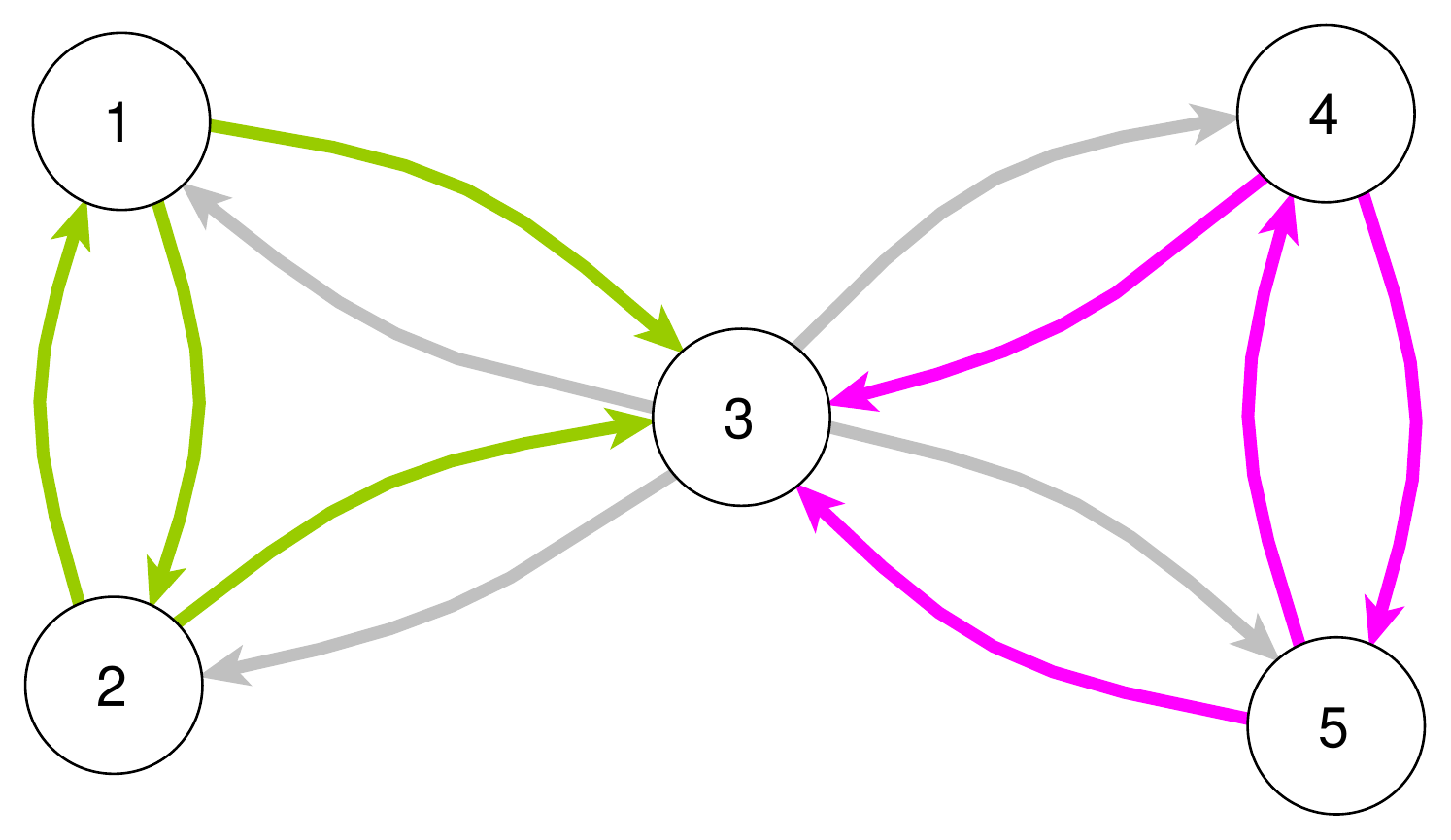}}
	\caption{\label{fig2} 
Illustration of the bow tie network studied in detail in the main text. When the process is Markovian and that the transitions between memory nodes are uniform, the second dominant left eigenvector, of eigenvalue $1/2$, is represented by the color code: 1 for green memory nodes, -1 for pink ones, and 0 for grey ones. The corresponding right eigenvector is obtained by symmetry.} 
\end{figure}
\begin{figure} 
	\centerline{\includegraphics[width=0.35\textwidth]{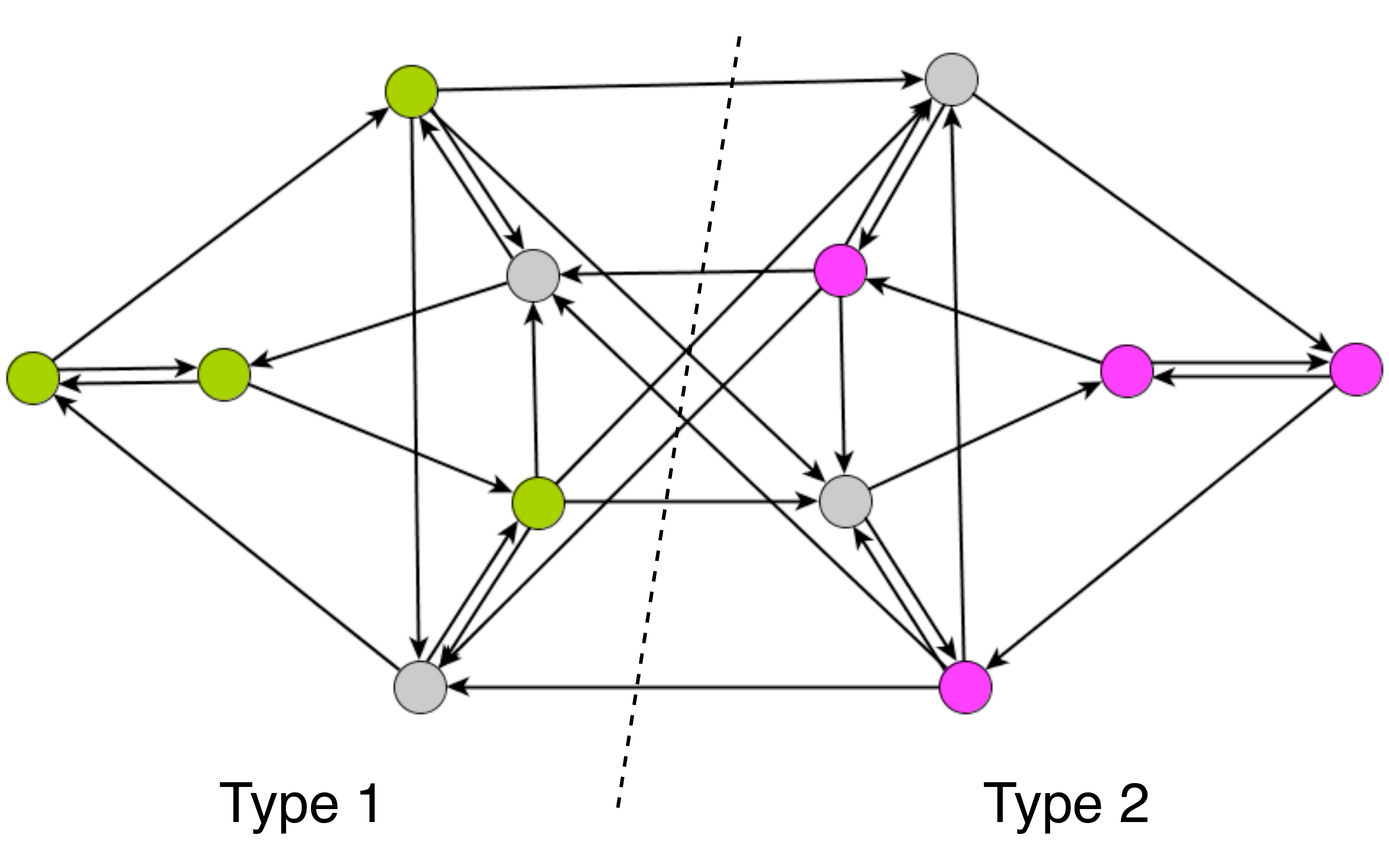}}
	\caption{\label{fig3} 
Representation of the memory network associated to Fig.~\ref{fig2}. The same color code has been used, and relative positions of memory nodes have been conserved. The non-Markovian process is defined by partitioning the network into two groups, and by assigning different types of transitions across and within groups. If a walker is located on a memory node on the left, its probability to jump to a neighbour in the left is weighted by a factor $1+\epsilon$. The factor is $1-\epsilon$ if the neighbour is on the right. } 
\end{figure}

\section{Numerical tests}

\begin{figure} 
	\centerline{\includegraphics[width=0.42\textwidth]{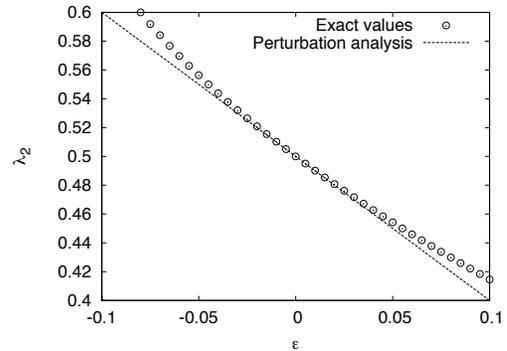}}
	\vspace{-0.7cm}
	\caption{\label{fig4} 
The validity of the theoretical prediction (13) for $\lambda_2$ is tested with the exact eigenvalue of the transition matrix of the dynamical process defined on Fig. \ref{fig2}. When $\epsilon=0$, the process is Markovian and $\lambda_2=1/2$. Increasing values of $\epsilon$ tend to slow down the relaxation towards stationarity.} 
\end{figure}

In order to test the validity of (\ref{res}), let us first focus on the toy example illustrated in Figure \ref{fig2}.
The physical network is composed of 5 nodes and 6 edges. The corresponding memory network, composed of 12 memory nodes, is illustrated in Figure \ref{fig3}.  The interplay between flow and topological structure is investigated by defining a non-Markovian process 
on the underlying topology. To do so, memory nodes are partitioned into two groups. Those on the left of the Fig. 3 are defined to be of type 1, and those on the right of type $2$. The biases associated to non-Markovianity are defined as follow:
the weight of a transition between nodes of the same type is $1+\epsilon$, while it is $1-\epsilon$ for different types. After a proper renormalization, one finds for instance $ T(\vv{12} \rightarrow \vv{21)}=1/2$, $ T(\vv{13} \rightarrow \vv{32})=(1+\epsilon)/4$, $ T(\vv{13} \rightarrow \vv{34})=(1-\epsilon)/4$, etc. By definition, the process is Markovian when $\epsilon=0$, and transitions between neighbouring memory nodes are uniform. In that case, the second dominant (left) eigenvector, represented in Figs. \ref{fig2} and \ref{fig3}, has an eigenvalue $1/2$, and is in line with the overlapping modular structure of the system.  Based on (\ref{res}), it is straightforward  to show that 
$ \Delta \lambda_2  = -\epsilon$, thereby
confirming that increasing intra-community flow tends to decrease the spectral gap, and thus to slow down relaxation towards stationarity. Increasing inter-community flow, that is flows that are orthogonal to the topological structure, instead accelerates diffusion. The accuracy of the perturbation analysis, and of the corresponding interpretation, is successfully compared to the exact eigenvalue $\lambda_2$ of the process (see Figure \ref{fig4}).

\begin{figure} 
	\centerline{\includegraphics[width=0.38\textwidth]{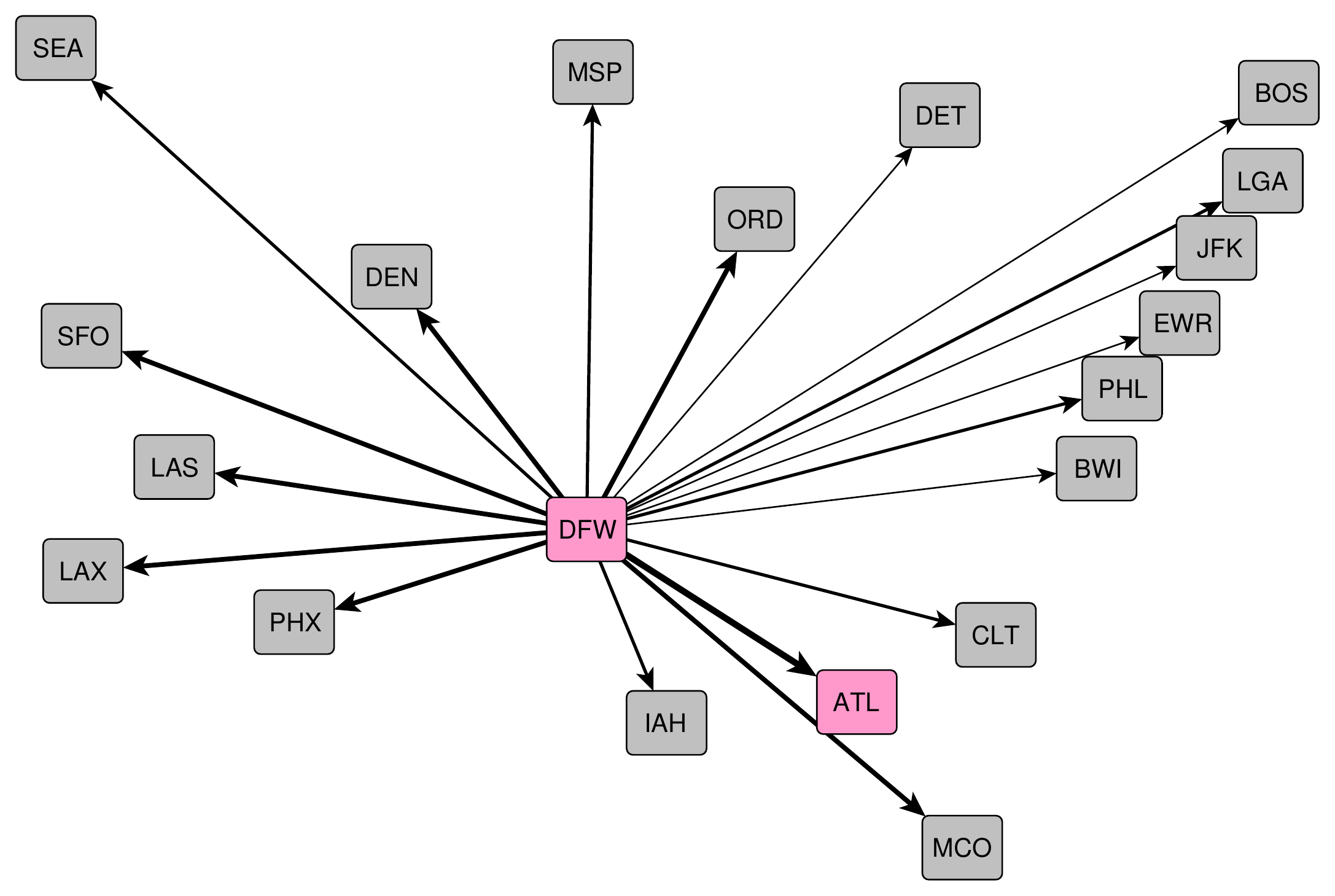}}
	\centerline{\includegraphics[width=0.38\textwidth]{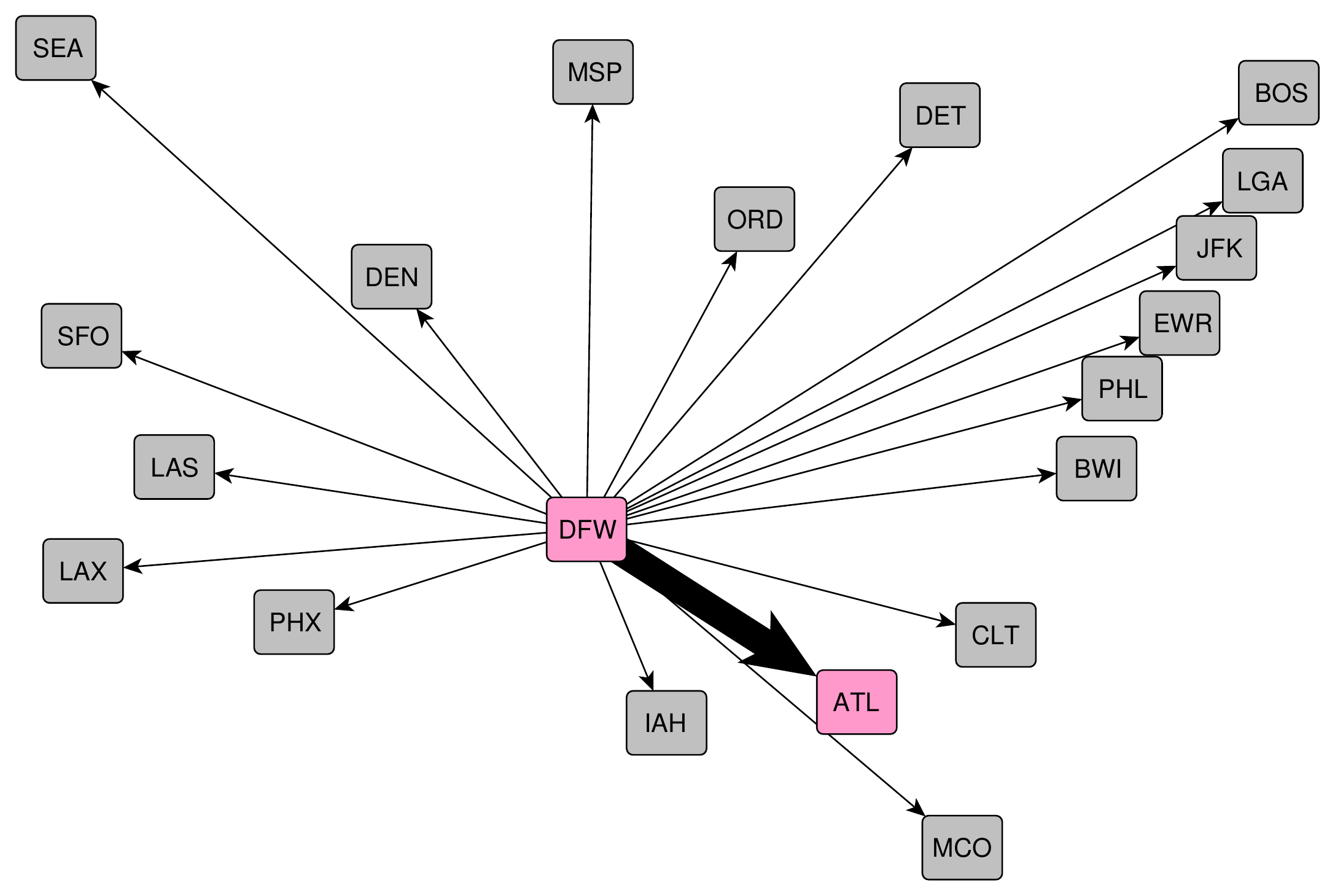}}
	\caption{\label{fig5} 
Illustration of the 19 flight legs leaving Dallas (DFW) after coming from Atlanta (ATL). In a first-order Markov model (upper), the probability to follow a link, represented by its width, is proportional to the total volume of traffic towards that airport, irrespectively of the origin of the passenger. In a second-order Markov model (lower), however, the probability to follow link depends on where the passengers come from, with a strong bias towards their origin.} 
\end{figure}

\begin{figure} 
	\centerline{\includegraphics[width=0.42\textwidth]{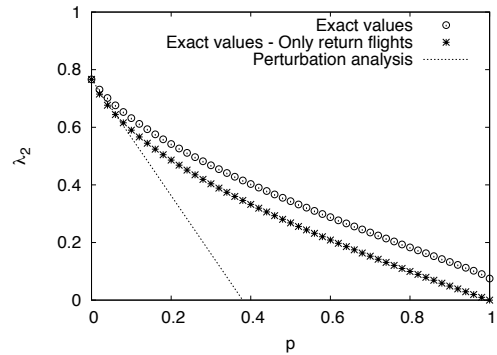}}
	\vspace{-0.7cm}
	\caption{\label{fig6} 
Effect of memory on $\lambda_2$ for the airport network. When $p=0$, the process is first-order Markov and when $p=1$, it is second-order Markov. Memory significantly slows down diffusion, with $\lambda_2$ going from $0.77$ to $0.07$ from one end to the other, with a mixing time going from $1.3$ to $14.28$, i.e. a slowndown by a factor of $11$. When only return flights are considered for the second-order Markov model, this slow-down is even more important, and the process becomes non-ergodic, with $\lambda_2=0$ when $p=1$.
} 
\end{figure}

As a second example, we consider airline pathways of passengers between US airports. The data-set, made publicly available by the Research and Innovative Technology Administration, contains detailed information about trip itineraries between 464 airports in the US. Modeled as a memory network, memory nodes $\vv{ij}$ represent flight legs from airports $i$ to $j$, and links represent connected flight legs, say $\alpha= \vv{ij}$ to $\beta=\vv{jk}$ (see Fig. \ref{fig5}). Links carry a weight proportional to the number of passengers going to $k$ from $j$ after they arrived from $i$. For more information on how to build second-order Markov processes from the empirical data, and a study of the effect of memory on modular structure, entropy rate and disease spreading, we refer to \cite{Rosvall} where the data was originally collected and compiled. In the following, we restrict the scope to a subset of the top 20 airports in terms of traffic, in order to ensure that the memory network is strongly connected, and that no teleportation is necessary in order to make the dynamics ergodic \cite{telep}. Because the airport network is weighted, in contrast with previous example, the transition matrix of the Markovian process is now given by 
\begin{equation}
\label{marW}
T^{M}_{\alpha \beta}= 
\frac{w_\beta}{\sum_\beta w_{\beta}},
\end{equation}
in order to provide an adequate baseline, and not uniform as in (\ref{mar}). $w_\beta$ is the total number of passengers along flight leg $\beta$, independently on where they come from.
In order to investigate the effect of memory, we consider a random process driven by the tunable transition matrix ${\rm T^{(p)}}= p {\rm T} + (1-p) {\rm T^M}$, that is a random walk process given by the first-order Markov process with probability $(1-p)$, and by the second-order Markov process with probability $p$. This hybrid process models the diffusion of an item, e.g. a virus or a bank note \cite{broc}, which travels with passengers and changes owner inside cities with probability $p$ \cite{Rosvall}. 
As shown in Fig~\ref{fig6}, the perturbation analysis provides a good approximation of $\lambda_2$ for small values of $p$, and shows that memory significantly slows down the spreading process. This slowdown is driven by the high return rate of the flights observed in Fig.~\ref{fig5}. To show so, we have calculated $\lambda_2$ for the transition matrix $p {\rm T^R} + (1-p) {\rm T^M}$, where $ {\rm T^R}$ models a second-order process where only return flights are permitted, i.e.
\begin{equation}
T^{R}_{\vv{ij} \beta}= 
\begin{cases}
1     & {\rm for}~~ \beta = \vv{ji},\cr
0      & {\rm otherwise}.
\end{cases} 
\end{equation}
As expected, this process slows down even further the random walk process and ultimately leads to non-ergodic dynamics when $p=1$, as the corresponding memory network becomes disconnected in that case.

\section{Conclusion}

In this paper, we have identified a mechanism by which non-random pathways on network either slow down or accelerate diffusion:
what matters, in the asymptotic regime, is the redistribution of flows by memory across the bi-modular structure of the system. Our theoretical results, tested on artificial and empirical data, find several applications. First, the study of real-world pathways on networks is expected to bloom in the next few years, as pathway information is becoming increasingly available and that network theory fails to capture important constraints on flow in complex systems. Our work predicts that a slow-down should be observed in a vast majority of empirical systems, as memory tends to enhance flows in dense substructures \cite{Rosvall}. The results also provide interesting ideas for community detection, in particular on how memory affects the detection of overlapping modules by link partitioning \cite{evans,ahn,boldi}. At a more fundamental level, they revisit the concept of small-worldness \cite{Watts}, by showing that a system, however small it is from a topological point of view, can be arbitrarily large or small from a dynamical poinf of view because of memory. In that direction, it would also be interesting to explore the criterion (\ref{res}) behaves for models of random walks with memory \cite{boldi,Rosvall}, and to study further the spectral properties of dynamics in memory networks.

In this work, we have focused on random walks, but a similar analysis could be developed for other types of dynamical models where memory is expected to be present, such as synchronization, for which the second dominant eigenvalue of the Laplacian plays a central role \cite{syn2}, or 
epidemic spreading, where the epidemic threshold is related to the largest eigenvalue of the adjacency matrix of the graph \cite{epidemic}. In opinion dynamics, history-dependent spreading processes are also non-Markovian. In particular, let us mention complex diffusion (in opposition with simplex diffusion), where redundancy promotes diffusion \cite{Centola}. Another important area of application  concerns multiplex networks \cite{Multi1,Multi1b,Multi2}, that is networks when nodes communicate through different interaction channels, and where information can be encapsulated inside one layer of communication, thereby leading to path-dependent transmissibility \cite{Min}. As an interesting research direction, let us note that diffusion on multiplex networks can be seen as a model for second-order Markov diffusion, where links of different types are partitioned into different layers in the multiplex.

Finally, this work sheds new light on a question that has puzzled researchers in recent years: does temporality accelerate or slow down diffusion in temporal networks \cite{Holme2012,Rocha2010,Karsai2011,Peruani2011,Masuda2013}. Previous research has shown that temporality affects diffusion  in different ways, because of the burstiness of the temporal patterns \cite{Lambiotte2013}, e.g. through the so-called waiting time paradox \cite{Miritello2011}, or of the temporal correlations between consecutive contacts present in the data \cite{Kovanen2011,Rocha2012}. In the latter case, the dynamics can be modeled as a non-Markovian process, since the next contact depends not only on the current one, but also on the previous ones \cite{Pfitzer} \footnote{When modelling the effect of non-exponential waiting time distributions on diffusion, the dynamics shows a different type of non-Markovianity, and the dynamics leads to integro-differential equations in time \cite{Hoffmann2012}. }. It has recently been shown that real-world temporal data are well-reproduced by second-order Markov models, and observed that the spectral gap may either increase or decrease due to memory  \cite{Scholtes}. Our work explicitly identifies the mechanism behind this change: the redistribution of flows compared to the Fiedler vector because of memory. 

\begin{acknowledgments} 

We would like to thank J.-C. Delvenne and I. Scholtes for fruitful discussions. We acknowledge financial support from F.R.S-FNRS and from the COST Action TD1210 KnowEscape. This paper presents research results of the Belgian Network DYSCO (Dynamical Systems, Control, and Optimization), funded by the Interuniversity Attraction Poles Programme, initiated by the Belgian State, Science Policy Office. 

\end{acknowledgments}

\end{document}